\documentclass[slovak,12pt,a4paper]{article}
\usepackage{t1enc}
\usepackage[english]{babel}
\usepackage[latin2]{inputenc}
\usepackage{graphicx}
\usepackage{amssymb,amsmath,babel,a4wide,epsfig,eufrak,amsfonts}
\usepackage{fancyhdr}
\usepackage{cite,./mcite}

\numberwithin{equation}{section} \numberwithin{figure}{section}
\numberwithin{table}{section}

\begin{document}

\clubpenalty=9999
\widowpenalty=9999

\vfill \vspace{2cm}
\begin{minipage}[m]{1.0\columnwidth}

\end{minipage}
\hfill{}
\begin{minipage}[m]{0.4\columnwidth}
\centering{DESY--08--063\\
CPHT--RR025.0408\\

\quad

\quad

\quad

\quad

\quad

\quad

 \Large \par}
\end{minipage}
\begin{center}
{\large{\bf{$Z$ and $W^\pm$ production associated with quark-antiquark pair\\
in $k_T$-factorization at the LHC}}}

\quad

\quad

\quad

Michal De\'ak$^1$, Florian Schwennsen$^{2}$

\quad

$^{1}$\textit{DESY, Hamburg, Notkestrasse 85, D--22607 Hamburg, Germany}\\
$^2${\it Centre de Physique Th\'eorique, \'Ecole Polytechnique, F--91128 Palaiseau, France }\\
\end{center}

\quad

\quad

\quad

\begin{abstract}
We calculate and analyze $Z$ and $W^{\pm}$ production in association with quark-anti\-quark pair in $k_T$-factorization. Numerical calculations are performed using the Monte Carlo generator {\sc Cascade} for proton proton collisions at LHC energy. We compare total and differential cross sections calculated in $k_T$-factorization approach with total differential cross sections obtained in LO and NLO calculations in collinear factorization approach. We provide strong evidence that some of  the effects of the NLO and even higher order collinear calculation are already included  in the LO $k_T$-factorization calculation.

\end{abstract}

\section{Introduction}

In the following years new discoveries are expected at the LHC
concerning physics within the Standard Model and beyond it.
The discovery of the Higgs boson and exclusion or affirmation of
possible extensions or alternatives to the Standard Model will be
of special interest. To be able to measure the proposed signals of
processes which open the access to new physics a very good
understanding of the detectors and their responses to produced
particles will be needed. An accurate calibration of particle
detectors could be achieved by using processes with well known
cross sections in which particles with well known properties
are produced. A calibration of LHC detectors using $W$ or $Z$
signals is proposed in several publications
\cite{W/Zcal1,*W/Zcal2}. Moreover, the $W$ or $Z$ production is
important because it plays a significant role in  background
processes connected to Higgs production. Another experimental
motivation is provided by the  possibility to measure the
luminosity via $Z$ boson production \cite{Dittmar:1997md}.

At the Tevatron collider $W/Z$ production takes place at a typical
$x=\sqrt{M_W^2/s}\approx 0.04$ and hence is dominated by
scattering of quarks.
Because of the much higher energy, proton scattering at LHC will
allow smaller proton energy fractions and will be dominated by
gluon scattering.

The $W$ mass provides a hard scale and allows a perturbative
calculation of the hard matrix element. The resummation of large
logarithms of the form $[\alpha_s\ln(\mu^2/\Lambda_{\rm
QCD}^2)]^n$ (where $\mu^2\sim M_W^2$, $\mu^2 \gg \Lambda_{\rm
QCD}^2$) can be performed in the framework of the
Dokshitzer-Gribov-Lipatov-Altarelli-Parisi (DGLAP) equation
\cite{DGLAP1,*DGLAP2,*DGLAP3,*DGLAP4}, leading to the collinear
factorization into conventional parton densities and a hard
scattering matrix element. While in the conventional collinear
approach the longitudinal momentum fraction is considered to be
dominant, such that the transverse momenta of the partons can be
neglected as well as their virtualities, at small $x$ the
transverse momenta entering the hard matrix element should become
relevant.

At the LHC the larger center of mass energy allows $W/Z$
production at even smaller $x$ such that the production of
particles will be dominated by gluon-gluon fusion. Moreover, in
this situation we have to deal with two different large scales
($s\gg \mu^2 \gg \Lambda_{\rm QCD}^2$) and logarithms of the  form
$[\alpha_s\ln(1/x)]^n$ arise which have to be resummed. This is
realized by the leading logarithmic (LL)
Balitsky-Fadin-Kuraev-Lipatov (BFKL) equation
\cite{BFKL1,*BFKL2,*BFKL3,*BFKL4} or the
Ciafaloni-Catani-Fiorani-Marchesini (CCFM) evolution equation
\cite{CCFM1,*CCFM2,*CCFM3,*CCFM4} which additionally resums terms
of the form  $[\alpha_s\ln(\mu^2/\Lambda_{\rm QCD}^2)]^n$ and
$[\alpha_s\ln(\mu^2/\Lambda_{\rm QCD}^2)\ln(1/x)]^n$.
Just as for DGLAP, it is possible to factorize the cross section
into a convolution of process-dependent hard matrix elements with
universal parton distributions. But as the virtualities and
transverse momenta are no longer ordered (as it is the case in
DGLAP evolution), the matrix elements have to be taken off-shell,
and the convolution has to be  made also over transverse momenta with the
so-called {\it unintegrated parton densities}. This factorization
scheme is called {\it $k_T$-factorization} \cite{CC1,CC2} or {\it
semi-hard approach} \cite{GL1,*GL2} and will be used in this work.

There is also the  notion of {\it transverse momentum dependent}
(TMD) parton distributions
~\cite{Collins:1981uk,*Ji:2004wu,*Ji:2004xq,*Collins:2004nx,*Bacchetta:2005pr}.
But although in these approaches the transverse momentum of the
parton is taken into account as well, this is only the case on the
side of the parton density. The matrix element is calculated with
incoming on-shell partons, and transversal momenta of the incoming
partons are neglected. It has been shown \cite{Collins:2007pr} that
factorization within this approach is violated beyond NLO. In case
of the $k_T$-factorization approach used in this work this is also
expected. Indeed, it is well known that in the BFKL approach
beyond NLO multiple gluon exchange in the $t$-channel has to be
taken into account.

In this paper we calculate and analyze $Z$ and $W$ production
associated with two quark jets provided by gluon-gluon fusion in
$k_T$-factorization. We assume quasi-multi-Regge-kinematics (QMRK)
where the cluster of $W/Z$ and the two quarks is well separated in
rapidity from the proton remnants while the kinematics within that
cluster is considered without any further assumption. In
particular, we take into account the mass of the quarks. In this
kinematic regime a gauge independent off-shell matrix element can
be extracted due to high energy factorization. A similar calculation has been done in \cite{Baranov:2007arx},
where the authors calculated photon (instead of $Z/W$) production
in the same framework. We calculated the matrix element independently and extended it to massive
gauge bosons.
In our work on massive gauge bosons production we especially focus on the predictions for LHC and compare with a
collinear factorization based calculation.

When this paper was in preparation, we learned about another group \cite{Baranov:2008rt} working on this process as well using the same theoretical approach, but laying more emphasis on confronting the theoretical predictions with experimental data and examining the role of quark contributions.

The paper is organized in the following way: In section 2 we
describe notation, kinematics of the process and the calculation
of the matrix element. In section 3 we present numerical results
obtained from a calculation using the Monte Carlo generator {\sc
Cascade} \cite{CASCADE1,*CASCADE2}, where the matrix element
squared was implemented.
In section 4 we summarize the results and offer conclusions.

\section{Kinematics of $Z/W$ production and calculation of the hard matrix element}

We label the 4-momenta of incoming hadrons with masses $m_A$ and
$m_B$ by $p_A^{\prime}$ and $p_B^{\prime}$, respectively. In the
center of mass system they can be expressed in terms of invariant
 light like vectors $p_A$ and $p_B$
\begin{align}
p_A^{\prime} =& p_A+\frac{m_A^2}{s}p_B, &
p_B^{\prime} =& p_B+\frac{m_B^2}{s}p_A \label{pprimes} .
\end{align}

In the case of protons at the LHC we have $m_A^2=m_B^2=m_p^2$
which satisfies the relation $\frac{m_p^2}{s}\ll 1$. Therefore, 
we can neglect the masses in Eqs. \eqref{pprimes}
and use $p_{A,B}$ instead of $p_{A,B}^{\prime}$.

\begin{figure}
\begin{center}
\epsfig{figure=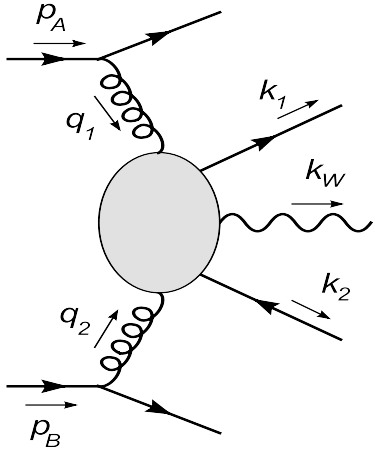,width=5cm,height=6cm,clip}
\caption{Labeling and flow of momenta of the process $pp\to
q\,(W/Z)\,\bar{q}\,X$.} \label{gen}
\end{center}
\end{figure}

It is convenient to use Sudakov decomposition for all momenta
present in the calculation (see also Fig.~\ref{gen}\footnote{These and the following diagrams were drawn in JaxoDraw~\cite{Binosi:2003yf}.}) by decomposing them into components
proportional to $p_A$ and $p_B$, and a remainder perpendicular to
both of them
\begin{equation}
k_i=\alpha_i p_A+\beta_i p_B +k_{i\perp},
\end{equation}
where $i\in\{1,2,W(Z)\}$ for outgoing particles, and
\begin{align}
q_1=& \alpha p_A+\beta_{q_1} p_B +q_{1\perp}, & q_2=&\alpha_{q_2}
p_A+\beta p_B +q_{2\perp}
\end{align}
for the gluons entering the hard matrix element. It is also
convenient to introduce Euclidean two dimensional vectors
$\vec{k}_{i}$ and $\vec{q}_{j}$ which satisfy the relations
$\vec{k}_{i}^{\,2}=-k_{i\perp}^2\geq 0$ and
$\vec{q}_{j}^{\,2}=-q_{j\perp}^2\geq 0$.

In QMRK we have
\begin{align}
\alpha \gg & \beta_{q_1}, & q_1^2=& -\vec{q}_{1}^{\,2}=t_1,\\
\beta \gg & \alpha_{q_2}, & q_2^2=& -\vec{q}_{2}^{\,2}=t_2, \\
 \alpha_i\beta_i=&\frac{m_i^2+\vec{k}_{i}^{\,2}}{s},\label{onshell}
\end{align}
where $i\in\{1,2,W (Z)\}$, and $m_i$ are the corresponding masses of
outgoing particles. The invariants  $t_1$ and $t_2$ describe the
momentum transfer between the cluster formed by the quarks and the
$W$ ($Z$) boson on one hand and the incoming protons on the other hand. Due to the strong ordering
in $\alpha$ and $\beta$ one can neglect terms proportional to
$\beta_{q_1}$ and $\alpha_{q_2}$ in the calculation.

It is useful to introduce a set of Mandelstam variables describing
the system
\begin{subequations}
\label{mandels}
\begin{align}
\hat{s}=&(q_1+q_2)^2= \alpha\beta s-(\vec{q}_{1}+\vec{q}_{2})^2,\\
\hat{s}_1=&(k_1+k_W)^2, & \hat{s}_2=&(k_2+k_W)^2,\\
\hat{t}_1=&(q_1-k_1)^2, & \hat{t}_2=&(q_2-k_2)^2,\\
\hat{u}_1=&(q_1-k_2)^2, & \hat{u}_2=&(q_2-k_1)^2,
\end{align}
\end{subequations}
related by
\begin{align}
&\hat{u}_1+\hat{t}_2+\hat{s}=t_1+t_2+m_2^2+\hat{s}_1, &
&\hat{u}_2+\hat{t}_1+\hat{s}=t_1+t_2+m_1^2+\hat{s}_2 .
\end{align}
It is  convenient to introduce transverse masses defined by
\begin{align}
\label{transm} m_{i\perp}=& \sqrt{m_i^2+\vec{k}_{i}^2}, &
m_{q\perp}=&\sqrt{\hat{s}+(\vec{q}_{1}+\vec{q}_{2})^2},
\end{align}
and longitudinal momentum fractions of the produced particles
$x_i=\frac{\alpha_i}{\alpha}$. Combining these relations with Eqs.
(\ref{onshell}, \ref{mandels}) one finds that -- in the end --
the matrix element of $W$ or $Z$ production associated with a quark-antiquark pair can be expressed in terms of independent Mandelstam variables
defined in Eqs. \eqref{mandels}, transverse masses and variables
$x_{1,2,W(Z)}$.

In the $k_T$-factorization formalism the hadronic and partonic
cross section are related as follows:
\begin{multline}\label{hcros}
  d\sigma(pp\to q \,(W/Z)\,\bar{q}\, X ) =
\int \frac{d\alpha}{\alpha}\int d\vec{q}_1^2\int\frac{d\phi_1}{2\pi}
\mathcal{A}(\alpha,\vec{q}_1^2,\mu^2) \\
\times\int \frac{d\beta}{\beta}\int
d\vec{q}_2^2\int\frac{d\phi_2}{2\pi}
\mathcal{A}(\beta,\vec{q}_2^2,\mu^2)
  d\hat\sigma(g^* g^* \to q \,(W/Z)\,\bar{q} ) ,
\end{multline}
where $\mathcal{A}$ is the unintegrated gluon density in a proton
and $\phi_{1,2}$ is the angle of $\vec{q}_{1,2}$ with respect to
some fixed axis in the azimuthal plane. The argument $\mu^2$ of unintegrated gluon densities is the factorization scale.
The partonic cross section is denoted by $ d\hat\sigma$.

\begin{figure}[htb]
\label{fullSet}
\begin{center}
\epsfig{figure=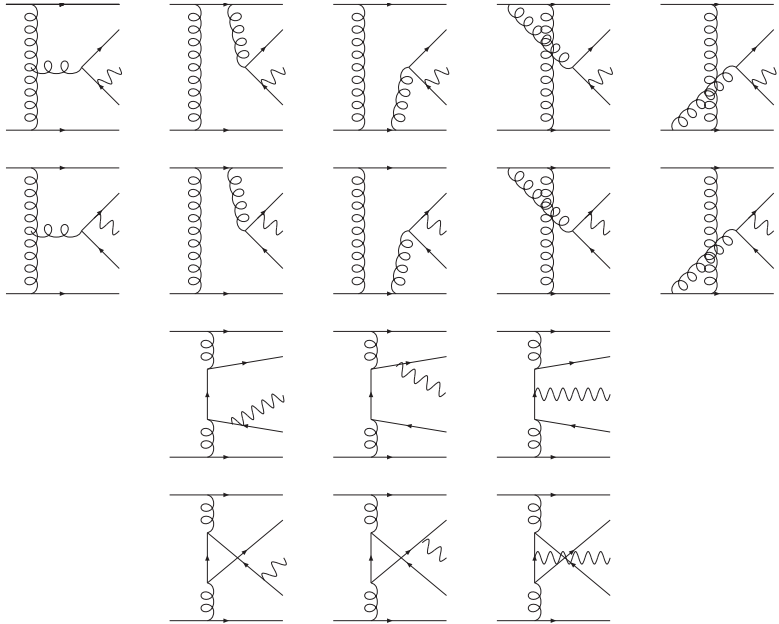,width=10.5cm,clip}
\caption{Full set of diagrams contributing to $W/Z$ production via off-shell
gluon-gluon fusion.} \label{fullDS}
\end{center}
\end{figure}

Since the incoming gluons of the matrix element entering this
partonic cross section are off-shell, the calculation differs from
that of a hard matrix element in the collinear approach
significantly. To guarantee gauge invariance, the process with
off-shell incoming particles has to be embedded into the
scattering of on-shell particles.  The extracted off-shell matrix element is of
course independent of the specific choice of the particles in
which the scattering process is embedded. Therefore, we replace
the protons by quarks for the calculation of the hard matrix
element. All diagrams for the discussed process are shown in Fig.~\ref{fullDS}.

\begin{figure}[htb]
\begin{center}
\epsfig{figure=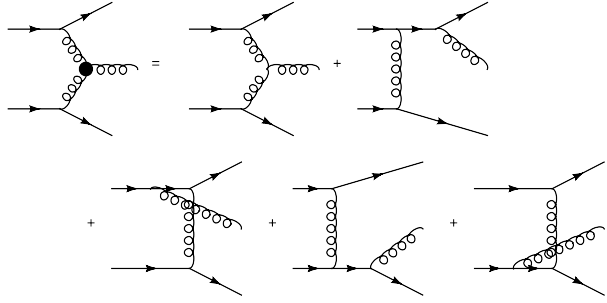,width=8cm,height=4cm,clip}
\caption{Diagrams contributing to the Lipatov vertex. }\label{effV}
\end{center}
\end{figure}

The first two rows of Fig.~\ref{fullDS}
include also  non-factorizing (`non-resonant') diagrams which factorize
only in the sum. To make this factorization apparent already at this level, one can sum up the different diagrams of one gluon production in quark-quark scattering leading to one effective diagram with an effective vertex (see Fig.~\ref{effV}). By working in Feynman gauge one obtains the well known  Lipatov vertex \cite{Lipatov:1976zz}:
\begin{equation}
\begin{aligned}
\Gamma_{\sigma\tau}^\nu(q_1,q_2)&=\frac{2p_{A\tau}p_{B\sigma}}{s}\Bigg(
  \frac{2t_1+m_{q\perp}^2}{\beta s}p_A^\nu
  -\frac{2t_2+m_{q\perp}^2}{\alpha s}p_B^\nu
-(q_{1\perp}-q_{2\perp})^\nu\Bigg).
\end{aligned}\label{eq:lipatovvertex}
\end{equation}
It can be shown that this vertex obeys the Ward identity. By this procedure, the first two rows of  Fig.~\ref{fullDS} are each replaced by just one diagram.

Strong ordering of Mandelstam variables $s$ and $t_{1,2}$ allows us to
make a simplification of the coupling of gluons to incoming
quarks. By neglecting the exchanged momentum in the vertex, we get
an eikonal vertex which does not depend on the spin of the particle
coupled to gluon and preserves its spin. In detail, it reads
\begin{equation}\label{eikver}
-i\bar{u}(\lambda_1^\prime,p_A-q_1)\gamma^\mu
u(\lambda_1,p_A)\quad\longrightarrow\quad
-2ip_A^\mu\delta_{\lambda_1^\prime,\lambda_1} .
\end{equation}
With the help of Eq.~\eqref{eikver} it is
possible to remove the external quark lines and attach so-called
`non-sense' polarizations to the incoming gluons:
\begin{align}
  \epsilon_{q_1}^\mu =& \frac{\sqrt{2}\,p_A^\mu}{\sqrt{s}} ,&
  \epsilon_{q_2}^\nu =& \frac{\sqrt{2}\,p_B^\nu}{\sqrt{s}} .
\end{align}

Instead of Feynman gauge, one can
choose an appropriate axial gauge \cite{CC1} $n\cdot A=0$
with the gauge vector
\begin{equation}
  \label{eq:gaugevector}
  n^\mu = ap_A^\mu+bp_B^\mu \quad\text{with } \;a,b \in \mathbb{C} .
\end{equation}
The contraction of the eikonal coupling \eqref{eikver} with the
gluon polarization tensor in this gauge
\begin{equation}
\label{eq:polarizationtensor} d_{\mu\nu}^{(n)}(q) =
-g_{\mu\nu}+\frac{n_\mu q_\nu+q_\mu n_\nu}{nq}- n^2\frac{q_\mu
q_\nu}{(nq)^2}
\end{equation}
then reads
\begin{align}
  p_A^\mu d_{\mu\nu}^{(n)}(q_1) =& \frac{q_{1\perp\nu}}{\alpha} , &
  p_B^\mu d_{\mu\nu}^{(n)}(q_2) =& \frac{q_{2\perp\nu}}{\beta} .
\label{eq:eikonaltimespolarization}
\end{align}
In such a physical gauge the `non-resonant' diagrams vanish
since the direct connection of two eikonal couplings gives $p_A^\mu
d_{\mu\nu}^{(n)}p_B^\nu=0$ (in other words: the Lipatov vertex is to be replaced by the usual three gluon vertex).

In the case of heavy quark production
the polarization sum for the $s$-channel gluon reduces to its
Feynman gauge analogue $-g_{\mu\nu}$ due to the heavy flavor
current conservation. The same simplification takes place in our calculation.
Nevertheless, we have to stress that in general the
polarization sum stays in its complex form.
Of course, both ways to calculate the matrix element are equivalent due to gauge invariance.

The sum over the physical polarizations $\eta$ of the $W$ boson reads
\begin{equation}\label{gauge}
\sum_\eta\epsilon^\mu(\eta,k_W)\epsilon^{\ast\nu}(\eta,k_W)
=-g^{\mu\nu}+\frac{k_W^\mu k_W^\nu}{m_W^2} .
\end{equation}
It is equivalent to replace the polarization sum by
\begin{equation}\label{gauge2}
\sum_\eta\epsilon^\mu(\eta,k_W)\epsilon^{\ast\nu}(\eta,k_W)
\quad\rightarrow\quad -g^{\mu\nu},
\end{equation}
and to add also the contribution of the Goldstone boson emission
diagrams, where the $W$ boson is replaced by a Goldstone boson
with mass $m_W$. This is in analogy of using the Feynman-t'Hooft
gauge instead of the unitary gauge. We have calculated the squared
matrix element in both ways as a crosscheck.

Expressions for the single diagrams in Fig.~\ref{fullDS}
-- where the first diagrams are already combined using the Lipatov vertex of Eq.~\eqref{eq:lipatovvertex} --
are listed here (the hat marks contraction with Dirac-matrices):
\begin{equation}
\begin{aligned}
&\mathcal{M}_{1\mu\nu}^{ab}=-ig_wg_s^2K_{W/Z}\;\bar{u}[t^b,t^a]\frac{\hat{\Gamma}_{\mu\nu}(q_1,q_2)}{\hat{s}}\frac{-\hat{k}_2-\hat{k}_W+m_1}{\hat{s}_2-m_1^2}\hat{\epsilon}(v_q-a_q\gamma^5)v,\\
&\mathcal{M}_{2\mu\nu}^{ab}=-ig_wg_s^2K_{W/Z}\;\bar{u}\hat{\epsilon}(v_q-a_q\gamma^5)
\frac{\hat{k}_1+\hat{k}_{W}+m_2}{\hat{s}_1-m_2^2}[t^b,t^a]\frac{\hat{\Gamma}_{\mu\nu}(q_1,q_2)}{\hat{s}}v,\\
&\mathcal{M}_{3\mu\nu}^{ab}=-ig_wg_s^2K_{W/Z}\;\bar{u}t^a\gamma_\mu\frac{\hat{k}_1-\hat{q}_1+m_1}{\hat{t}_1-m_1^2}t^b\gamma_\nu\frac{-\hat{k}_2-\hat{k}_W+m_1}{\hat{s}_2-m_1^2}\hat{\epsilon}(v_q-a_q\gamma^5)v,\\
&\mathcal{M}_{4\mu\nu}^{ab}=-ig_wg_s^2K_{W/Z}\;\bar{u}\hat{\epsilon}(v_q-a_q\gamma^5)\frac{\hat{k}_1+\hat{k}_W+m_2}{\hat{s}_1-m_2^2}t^a\gamma_\mu\frac{\hat{q}_2-\hat{k}_2+m_2}{\hat{t}_2-m_2^2}t^b\gamma_\nu v,\\
&\mathcal{M}_{5\mu\nu}^{ab}=-ig_wg_s^2K_{W/Z}\;\bar{u} t^a\gamma_\mu\frac{\hat{k}_1-\hat{q}_1+m_1}{\hat{t}_1-m_1^2}\hat{\epsilon}(v_q-a_q\gamma^5)\frac{\hat{q}_2-\hat{k}_2+m_2}{\hat{t}_2-m_2^2} t^b\gamma_\nu v,\\
&\mathcal{M}_{6\mu\nu}^{ab}=-ig_wg_s^2K_{W/Z}\;\bar{u}\hat{\epsilon}(v_q-a_q\gamma^5)\frac{\hat{k}_1+\hat{k}_W+m_2}{\hat{s}_1-m_2^2} t^b\gamma_\nu\frac{\hat{q}_1-\hat{k}_2+m_2}{\hat{u}_1-m_2^2} t^a\gamma_\mu v,\\
&\mathcal{M}_{7\mu\nu}^{ab}=-ig_wg_s^2K_{W/Z}\;\bar{u} t^b\gamma_\nu\frac{\hat{k}_1-\hat{q}_2+m_1}{\hat{u}_2-m_1^2} t^a\gamma_\mu\frac{-\hat{k}_2-\hat{k}_W+m_1}{\hat{s}_2-m_1^2}\hat{\epsilon}(v_q-a_q\gamma^5) v,\\
&\mathcal{M}_{8\mu\nu}^{ab}=-ig_wg_s^2K_{W/Z}\;\bar{u} t^b\gamma_\nu\frac{\hat{k}_1-\hat{q}_2+m_1}{\hat{u}_2-m_1^2}\hat{\epsilon}(v_q-a_q\gamma^5)\frac{\hat{q}_1-\hat{k}_2+m_2}{\hat{u}_1-m_2^2} t^a\gamma_\mu  v,\\
\end{aligned}
\label{eq:matrixelements}
\end{equation}

with the short hand notations $\bar{u}\equiv \bar{u}(\lambda,k_1)$, $v\equiv v(\lambda^\prime,k_2)$, $\hat{\epsilon}\equiv\hat{\epsilon}(\eta,k_W)$, and where $\eta$, $\lambda$ and $\lambda^\prime$ label the helicity/
spins of the corresponding particles. Color factors are
represented by Gell-Mann matrices $t^a$, $t^b$. The factors $v_q$, $a_q$
and $K_{W/Z}$ encode the $W$ and $Z$ coupling. For $W$ boson we
have $v_q=a_q=1$ and $K_{W}=V_{ud}\frac{1}{2\sqrt{2}}$, where
$V_{ud}$ is the corresponding element of Cabibbo-Kobayashi-Maskawa
matrix. For $Z$ we have $a_u=\frac{1}{2}$,
$v_u=\frac{1}{2}-\frac{4}{3}\sin^2\theta_W$ and $a_d=-\frac{1}{2}$,
$v_d=-\frac{1}{2}+\frac{2}{3}\sin^2\theta_W$ and
$K_{Z}=\frac{1}{2\cos\theta_W}$, where $\theta_W$ is the
Weinberg angle. In the latter case $m_1$ equals $m_2$, and $m_W$ is replaced by $m_Z$.

If we make use of the Eq.~\eqref{gauge2} to replace the
polarization sum,  one has to add diagrams and corresponding
amplitudes with Goldstone bosons with couplings
\begin{equation}\label{Gtoq}
-ig_{w}K_{W/Z}\Big(\frac{m_2-m_1}{m_{W/Z}}\,v_q-\frac{m_1+m_2}{m_{W/Z}}\,a_q\gamma^5\Big) .
\end{equation}

Finally, the square of the amplitude averaged over initial helicities and
colors of gluons and summed over spins/ helicities and colors of
final particles can be written as
\begin{equation}\label{MEsquared}
\frac{1}{4}\frac{1}{(N_c^2-1)^2}|\mathcal{M}|^2=\frac{1}{4}\frac{1}{(N_c^2-1)^2}\sum_{\lambda,\lambda^\prime,
\eta,a,b}{\rm Tr}_{\rm color}\Bigg\{\Bigg|\sum_{i=1}^8
\epsilon_{q_1}^\mu\epsilon_{q_2}^\nu
\mathcal{M}_{i\mu\nu}^{ab}\Bigg|^2\Bigg\}.
\end{equation}

By evaluating the traces over the products of Gell-Mann color
matrices, one encounters two possible cases of color factors
\begin{align}
 {\rm Tr}\{t^at^bt^at^b\}=&-\frac{1}{4}\frac{N_c^2-1}{N_c},&
 {\rm Tr}\{t^at^bt^bt^a\}=&\frac{1}{4}\frac{(N_c^2-1)^2}{N_c},
\end{align}
where $N_c=3$ is the number of colors.

Finally, the expression for the partonic off-shell cross section
appearing in  Eq.~\eqref{hcros} to calculate the hadronic
cross section is
\begin{equation}
\begin{aligned}
d\hat{\sigma}
(g^* g^* \to q \,(W/Z)\,\bar{q}\, )=&(2\pi)^{4}\delta^{(4)}\left(q_1+q_2-k_1-k_2-k_{W/Z}\right)\times\\
&\times\frac{1}{2\alpha\beta s} \frac{\alpha^2\beta^2
s^2}{t_1t_2}\frac{1}{4}\frac{1}{(N_c^2-1)^2}|\mathcal{M}|^2
\prod_{i\in\{1,2,W(Z)\}}\frac{d^3k_i}{(2\pi)^32E(k_i)} .
\end{aligned}
\end{equation}

The origin of the specific form of the flux factor and prefactor
$\frac{\alpha^2\beta^2 s^2}{t_1t_2}$ is formulated in
~\cite{CC1,CC2}. We summarize the most relevant aspects here. An
important feature of the whole calculation is that it is possible
to recover the result obtained in collinear factorization by
neglecting the transverse momenta of the gluons when they enter
the hard matrix element and instead integrate over them only in
the gluon densities. Due to factorization it is possible to keep
this connection not only for the full cross section, but also for
gluon densities and hard matrix element separately as well,
provided that the explicit manifestations of the factorization
formulae are phrased.

The key point is the observation that
\begin{equation}\label{eq:azav}
  \Big\langle 2\frac{q_{1\perp\mu}q_{1\perp\nu}}{q^2_{1\perp}}\Big\rangle_{\phi_1}
= -g^{\perp}_{\mu\nu} = \Big\langle
2\frac{q_{2\perp\mu}q_{2\perp\nu}}{q^2_{2\perp}}\Big\rangle_{\phi_2}
.
\end{equation}
As shown in
Eqs.~(\ref{eq:gaugevector}-\ref{eq:eikonaltimespolarization}), in
an appropriate gauge the polarization sum
$\frac{2p_{A\mu}p_{B\nu}}{s}$ can be replaced by
$\frac{2q_{1\perp\mu}q_{2\perp\nu}}{\alpha\beta s}$. Since in this
gauge one has to deal with exactly the same diagrams as in the
on-shell calculation, by dressing the off-shell matrix element
squared with the prefactor $\frac{\alpha^2\beta^2 s^2}{t_1t_2}$
and performing the averaging over azimuthal angles of the
`incoming' gluons, followed by taking the limit $t_1,t_2\to 0$,
one gets the collinear limit of the  matrix element squared. The
flux factor for off-shell gluons is defined as for on-shell
gluons with $\frac{1}{2\alpha\beta s}$. As the matrix element is
gauge invariant, this connection remains valid when one performs the
current calculation in a different gauge.

Due to the off-shellness of the incoming gluons and the three particle final state the final result of the matrix element squared is rather lengthy. 
For that reason, we calculated it independently and in different ways. One calculation followed directly the derivation above using Feynman gauge for 
the gluons, and has been performed using {\sc Mathematica}. A second calculation written in {\sc Form} \cite{Vermaseren:2000nd,Vermaseren:2002rp} used 
an axial gauge as described above such that the Lipatov vertices in \eqref{eq:matrixelements} are to be replaced by standard 
three-gluon-vertices. Moreover this second method used the  method of
orthogonal amplitudes, described in \cite{Baranov:2004pa}, which affects the fermionic part of the matrix element and with
which one is able to treat the matrix element squared in a more
compact way.\footnote{We also have  cross-checked numerically our results for the case of a produced photon instead of a $W/Z$ boson with those of the authors of \cite{Baranov:2007arx} whose cooperation we gratefully acknowledge.}

For this second method a few technical details are elaborated 
in the remainder of this section. The method of
orthogonal amplitudes is based on expressing a generic amplitude
$\widetilde{\mathcal{M}}$ (with one quark line) in terms of a set
of four independent operators $\hat{O}_i,\;i\in\{1,..,4\}$, which
satisfy orthogonality relations ${\rm
Tr}\{\hat{O}_i(\hat{k}_2-m_2)\overline{\hat{O}_j}(\hat{k}_1+m_1)\}=\|\hat{O}_i\|^2\delta_{ij}$
for any possible $i$ and $j$, where $\|\hat{O}_i\|$ is the
``norm'' of the operator $\hat{O}_i$. The projection of
$\widetilde{\mathcal{M}}$  by an operator $\hat{O}_i$ is performed
in the following way
\begin{equation}\label{projections}
\widetilde
{\mathcal{M}}^i=\frac{1}{\|\hat{O}_i\|}\sum_{\lambda,\lambda^\prime}\widetilde{\mathcal{M}}\,\bar{v}(\lambda^\prime,k_2)\overline{\hat{O}_i}
u(\lambda,k_1).
\end{equation}
The matrix element squared then has the following form
\begin{equation}
\sum_{\lambda,\lambda^\prime}|\widetilde{\mathcal{M}}|^2=\sum_{i}{|\widetilde{\mathcal{M}}^i|^2}.\label{eq:sumprojections}
\end{equation}
In our case the matrix element consists of up to five
Dirac-matrices (neglecting $\gamma^5$), after squaring one has to
evaluate traces of up to twelve of them. In contrast the method of
orthogonal amplitudes leads only to traces of up to eight
Dirac-matrices.

If one wants to consider also the $Z$ or $W^\pm$ coupling in the
Feynman diagram, one encounters a technical problem connected with
the appearance of the Dirac-matrix $\gamma^5$ in the expression
for the amplitude, leading to  terms which include Levi-Civita
tensors which later cancel. To avoid this complication, one can
split the expression for the amplitude into two parts, one which
does not include $\gamma^5$ and the other one which does  (to
separate the vector and axial part of the $Z$ or $W$ boson
coupling). For the part with $\gamma^5$ one uses a base of
operators $\hat{O}_i\gamma^5$. It is easy to check that they
satisfy the same orthogonality relation like the operators
$\hat{O}_i$. One also easily see that projections of amplitudes
in which $\gamma^5$ occurs do not contain terms with Levi-Civita
tensors. In doing so, we extend the method of  orthogonal amplitudes in a natural way.

Another complication comes from the presence of color factors in
the expressions which are not numbers but matrices. To treat the
projections as numbers, it is necessary to  separate the Feynman
diagrams into three groups according to different color factors,
namely
\begin{equation}
\begin{aligned}
C_1^{ab}&=t^at^b-t^bt^a ,\\
C_2^{ab}&=t^at^b ,\\
C_3^{ab}&=t^bt^a ,
\end{aligned}
\end{equation}
which form a vector $C^{ab}=(C^{ab}_1,C^{ab}_2,C^{ab}_3)$
(components of $C^{ab}$ are color factors of
$\mathcal{M}_{(1,2)\mu\nu}^{ab}$, $\mathcal{M}_{(3-5)\mu\nu}^{ab}$
and $\mathcal{M}_{(6-8)\mu\nu}^{ab}$ correspondingly). One can
then build a corresponding vector containing the sums of Feynman
diagrams without the color factors
$\mathcal{F}=(\mathcal{F}_1,\mathcal{F}_2,\mathcal{F}_3)$ such
that
\begin{equation}
\mathcal{M}^{ab}=(C^{ab})^T\mathcal{F} .
\end{equation}

The Lorentz indices have been dropped for simplicity. Using the matrix
\begin{equation}
\mathcal{C}_{ij}={\rm Tr}\{C_i^{ab}C_j^{ba}\} ,
\end{equation}
the expression for the square of the
matrix element takes the form
\begin{equation}
|\mathcal{M}|^2=\mathcal{F}^\dagger \mathcal{C}\mathcal{F},
\end{equation}
where combinations of $\mathcal{F}_i$ and $\mathcal{F}_j^*$ are calculated using
the projection method introduced in Eqs.~(\ref{projections}, \ref{eq:sumprojections}).
For the final simplification we have diagonalized the matrix
$\mathcal{C}$. After diagonalization of the matrix $\mathcal{C}$ only two diagonal elements remain nonzero.
This is expected because the quarks in the final state, in this process, can occur only in two possible color
states.

\section{Numerical studies}\label{sec:numstudies}

The last missing pieces needed to calculate the hadronic cross
section using Eq.~\eqref{hcros}, are the unintegrated gluon
densities. As mentioned in the introduction, there are two
equations suited to describe the evolution of an unintegrated
gluon density, namely BFKL \cite{BFKL1,*BFKL2,*BFKL3,*BFKL4}  and
CCFM \cite{CCFM1,*CCFM2,*CCFM3,*CCFM4}, respectively. Both have
been shown to agree on the leading logarithms in small {$x$}
\cite{Forshaw:1998uq,*Webber:1998we,*Salam:1999ft}, but the CCFM
evolution is valid in the domain of larger $x$ as well and,
moreover, matches in this region with DGLAP.
Therefore, we base our numerical studies on an unintegrated gluon density
obeying the CCFM equation, which has been implemented in the Monte
Carlo generator {\sc Cascade} \cite{CASCADE1,*CASCADE2}. We also investigate how the results change when using uPDFs generated by a different
procedure known as  KMR \cite{Kimber:2001sc}.

For this purpose, we implemented the matrix element squared as described above into {\sc Cascade}.
This implementation will be available in the next version of {\sc Cascade}.

We have used the unintegrated parton distribution function (uPDF) CCFM 2003 set 3 for the numerical
calculation.

To investigate the calculated matrix element as accurately as
possible, we neglect in this first study the effect of
hadronization of the final state.
We study in detail rapidity and transverse momentum distributions of the
produced gauge boson, quark and antiquark which (if one assumes
that quarks approximately determine jets) are the most important
observables in the experiment.

Furthermore, we compare the $k_T$-factorization approach
to the collinear one. For this purpose, we compare the
distributions obtained by our transverse momenta
dependent matrix element with distributions obtained from the
Monte Carlo generator {\sc Mcfm} \cite{MCFM,*Z+2jet1,*Z+2jet2}
which provides a calculation of the same process in the collinear
limit. In that case the transverse momenta coming from the
evolution are neglected. We also investigate in Sec.~\ref{uPDFs}
how the variation of unintegrated parton densities affect the
azimuthal angle and transverse momenta distributions.

As an artefact of the perturbative calculation, the results depend
on the renormalization scale $\mu_R$ and the factorization scale
$\mu_f$. In the CCFM formalism the hardest scale is set by the
emission angle of the hardest subcollision. It can be expressed in
terms of the energy of the subcollision
$\mu_f=\sqrt{\hat{s}+(\vec{q}_1+\vec{q}_2)^2}$. For the comparison
with collinear factorization calculations we have used as
renormalization scale $\mu_R=m_Z$ in $k_T$-factorization calculation
and in collinear calculation as well. We have also investigated other
possible choices (see subsection~\ref{uPDFs}).

\subsection{Comparison with LO collinear calculation}\label{Res2}

Our calculation of the hard matrix elements includes $W^{\pm}$ and
$Z$ production in association with all possible quark-antiquark
channels in gluon gluon fusion. Since the basic structure of all
these matrix elements is very similar, we present results only for
the typical case of $Zb\bar{b}$ production at LHC energies of
$\sqrt{s}=14{\rm TeV}$. The mass of the $b$-quark used is
$m_b=4.62~{\rm GeV}$.
For the collinear factorization calculations we use
the parton densities CTEQ6L1 \cite{CTEQ}.

\begin{figure}
\begin{center}
\epsfig{figure=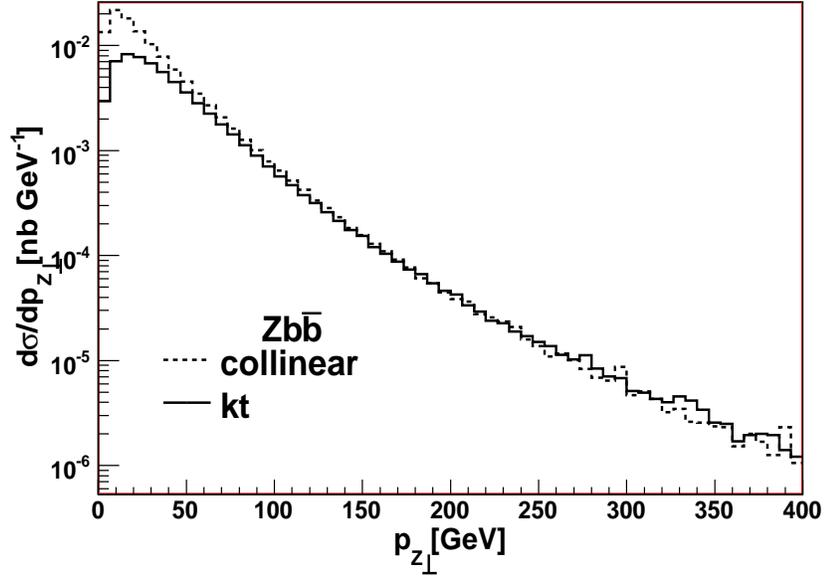,width=12cm,height=8.5cm,clip}
\caption{Transverse momentum distributions of the
produced $Z$ gauge bosons. Calculation with massive $b$-quarks. Both
calculations are in LO of perturbation series.} \label{histo Z
pt2}\label{histoZpt2}
\end{center}
\end{figure}

\begin{figure}
\begin{center}
\epsfig{figure=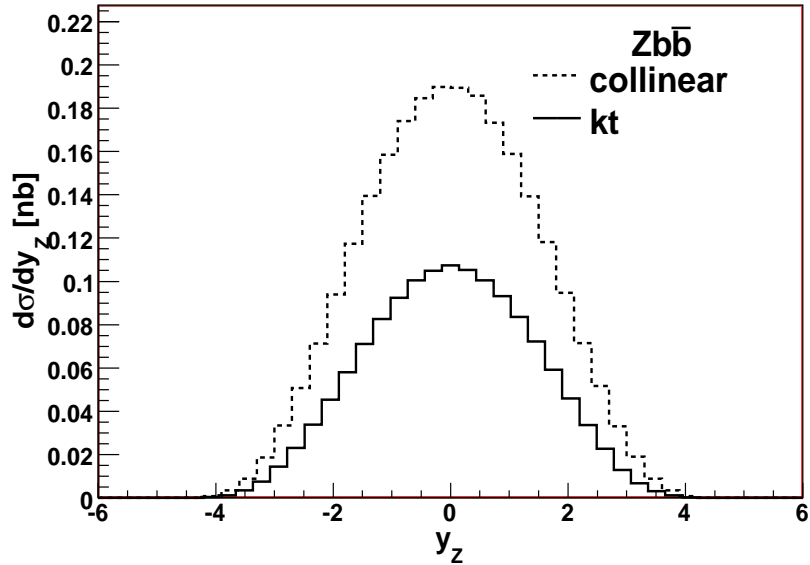,width=12cm,height=8.5cm,clip}
\caption{Rapidity distribution of the produced $Z$ gauge bosons.
Calculation with massive $b$-quarks.
Both calculations are in LO of perturbation series.} \label{histo
Z y}\label{histoZyLO}
\end{center}
\end{figure}

The total cross sections are comparable in magnitude, though they
differ considerably: $0.406\,{\rm nb}$ in $k_T$-factorization and
$0.748\,{\rm nb}$ in collinear factorization. The difference of
total cross sections stems from the different behavior at low
transversal momenta of final state particles (discussed later in
this section) where contributions from transversal momenta of the
initial state gluons play a significant role. It can be seen
that by applying a cut on the transversal momentum of 
the $Z$ boson $p_{Z\perp}>50\,{\rm GeV}$ the difference of the total
cross sections becomes smaller. With this additional cut one obtains cross sections of $0.118\,{\rm nb}$ in $k_T$-factorization and 
$0.141\,{\rm nb}$ in LO collinear calculation.

The total cross sections for other final states of interest are
given in Tab.~\ref{tab:totalcrosssections}.
\begin{table}
\begin{center}
\begin{tabular}{|c||c|c|c|c|}
\hline
final state   & $Zc\bar{c}$ & $Zb\bar{b}$ & $Zt\bar{t}$ & $W^+ s\bar{c}$, $W^- c\bar{s}$ \\
\hline
$\sigma_{\rm tot}$ [nb] & $0.430$  & $0.406$ & $0.525\cdot10^{-3 }$ & $1.92$ \\
\hline
\end{tabular}
\end{center}
\caption{Total cross sections for different final states, calculated in $k_T$-factorization using {\sc Cascade}.}\label{tab:totalcrosssections}
\end{table}

The transverse momentum and rapidity distributions of the vector
boson are shown in Fig.~\ref{histoZpt2} and \ref{histoZyLO},
respectively.
The comparison of the $k_T$-factorization approach to the
collinear shows that they agree in transversal
momentum distributions of $Z$ at high values of this quantity.
This is no surprise, since at high $p_{Z\perp}$ the contribution from initial state gluon
transverse momenta is expected to become small.

The rapidity distributions of the $Z$  show a similar behavior,
except for the overall normalization (Fig.~\ref{histoZyLO}).

\begin{figure}
\begin{center}
\epsfig{figure=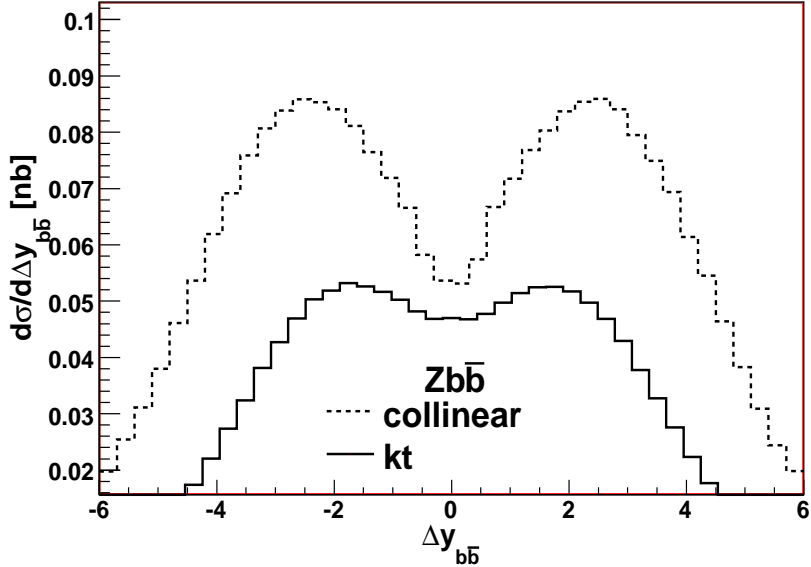,width=12cm,height=8.5cm,clip}
\caption{Distributions of the rapidity distance between quark and antiquark.
Calculation with massive $b$-quarks.
Both calculations are in LO of perturbation series.} \label{fig:rapdist}
\end{center}
\end{figure}

\begin{figure}
\begin{center}
\epsfig{figure=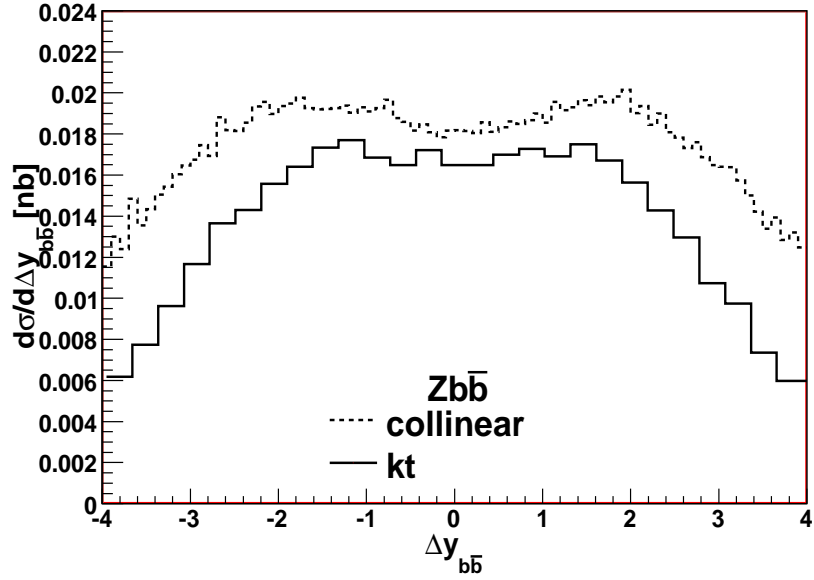,width=12cm,height=8.5cm,clip}
\caption{Distributions of the rapidity distance between quark and antiquark.
Calculation with massive $b$-quarks. A cut on $p_{Z\perp}>50\,{\rm GeV}$ has been applied.} \label{fig:rapdistShallower}
\end{center}
\end{figure}

To elaborate the difference between $k_T$- and collinear factorization,
we investigate more exclusive observables, like
the cross section differential in rapidity distance between quark and antiquark (Fig.~\ref{fig:rapdist}).
Both calculations show a two peak structure with a minimum at zero
rapidity, but the $k_T$-factorization result has a  considerably shallower minimum. 
The minimum in the case of
the collinear calculation gets shallower -- bringing together both calculations  -- when one again applies a cut on $p_{Z\perp}>50\,{\rm GeV}$ as one can see in 
Fig.~\ref{fig:rapdistShallower}.

\begin{figure}
\begin{center}
\epsfig{figure=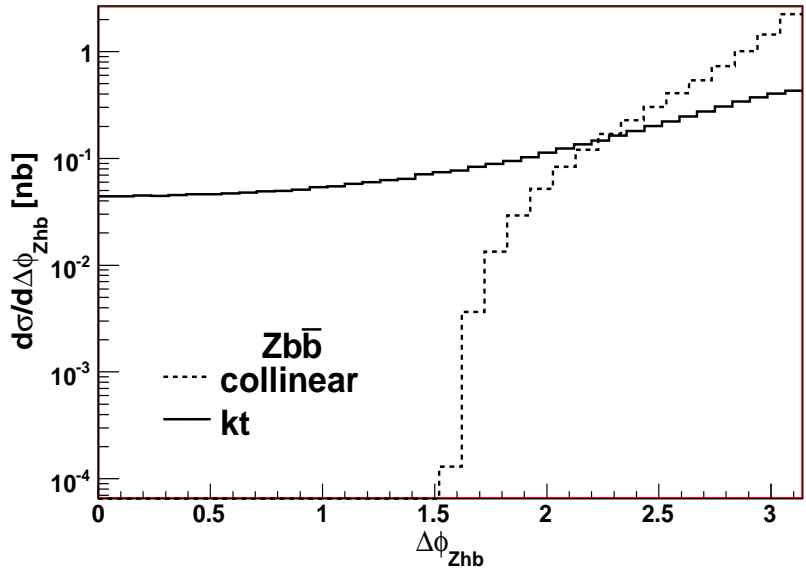,width=12cm,height=8.5cm,clip}
\caption{Distributions of the distance in azimuthal angle of $Z$ and
highest $p_\perp$ quark or antiquark.
Calculation with massive $b$-quarks.
Both calculations are in LO of perturbation series.}
\label{fig:angdistZq}
\end{center}
\end{figure}

In the distribution of the azimuthal angular distance
of $Z$ and  $\max(p_{b,\perp},p_{\bar{b},\perp})$ (Fig.~\ref{fig:angdistZq}) we observe that the region from
$0$ to $\pi/2$ is forbidden within the collinear calculation due to momentum conservation, which is not the case for $k_T$-factorization. This is caused by the contribution
from initial state gluon transversal momentum which allows the
transversal momenta of $Z$, $b$ and $\bar{b}$  to be unbalanced.
A larger spread of possible configurations causes that the
distribution in the $k_T$-factorization calculation flattens.

\subsection{Comparison with NLO collinear calculation}\label{Res1}

In collinear factorization the physical effect of the intrinsic transverse momenta of the initial gluons can not be described until higher order corrections are taken into account. Then additional real emissions lead to off-shell gluons and their transverse momenta.
Therefore, the significant differences between a calculation in the collinear factorization framework
 and $k_T$-factorization framework shown in the previous section encourage us to compare our LO calculation in
$k_T$-fac\-tor\-iza\-tion with a  NLO collinear calculation, since
CCFM evolution includes the high-energy part of the NLO collinear corrections. Since there are two off-shell initial gluons in a $k_T$-factorized $pp$-collision, one could
even call for a higher order collinear calculation to compare\footnote{Although we argue that already the LO $k_T$-factorization calculation includes in some sense higher order corrections,
one might ask for an extension to NLO.
So far
$k_T$-factorization based on CCFM evolution has been formulated
only at LO. On the other hand, since the BFKL equation has been
calculated at NLO accuracy \cite{Fadin:1998py,*Ciafaloni:1998gs},
in the small $x$ regime $k_T$-factorization can be formulated at
NLO accuracy as well \cite{Bartels:2006hg}. Nevertheless, an
implementation into a Monte Carlo generator is still outstanding.
Moreover, the calculation of an off-shell $2\to 3$ process at one
loop order is far beyond the scope of this work.}.

To compare with a collinear NLO calculation,
we use again the Monte Carlo generator {\sc Mcfm}. This Monte
Carlo generator provides the process $gg \to Zb\bar{b}$ at NLO only in
the massless quark limit. To avoid divergences, additional cuts
are applied on transversal momenta of quarks, on the invariant
mass of the $b\bar{b}$ pair, and on transversal momenta of a gluon
which is produced in diagrams of real NLO corrections. Transversal
momenta of produced quark, antiquark and gluon have to satisfy the
condition $p_\perp>4.62{\rm GeV}$ (corresponding to the mass of the $b$-quark). These
cuts on quark (antiquark) momenta are automatically applied in
{\sc Mcfm} when one is performing a calculation involving massless
quarks (antiquarks). We choose the parton density functions set
CTEQ6M \cite{CTEQ}.
The same cuts on transversal momenta of quark and
antiquark are then applied in {\sc Cascade} as well.

For the total cross sections, we obtain in the NLO collinear
factorization calculation $1.04\,{\rm nb}$, and in the
$k_T$-factorization calculation $0.429\,{\rm nb}$. The
difference of the total cross sections in $k_T$-factorization
calculation and the NLO calculation in collinear factorization
is of the same origin as the difference between the total
cross sections in section \ref{Res2} where comparison of
$k_T$-factorization calculation and NLO calculation in collinear
factorization is discussed. This is again illustrated by a cut on $p_{Z\perp}>50\,{\rm GeV}$ diminishing the difference between the cross sections ($0.125\,{\rm nb}$ for the $k_T$-factorization calculation and $0.165\,{\rm nb}$ for the NLO calculation in collinear factorization).

\begin{figure}
\begin{center}
\epsfig{figure=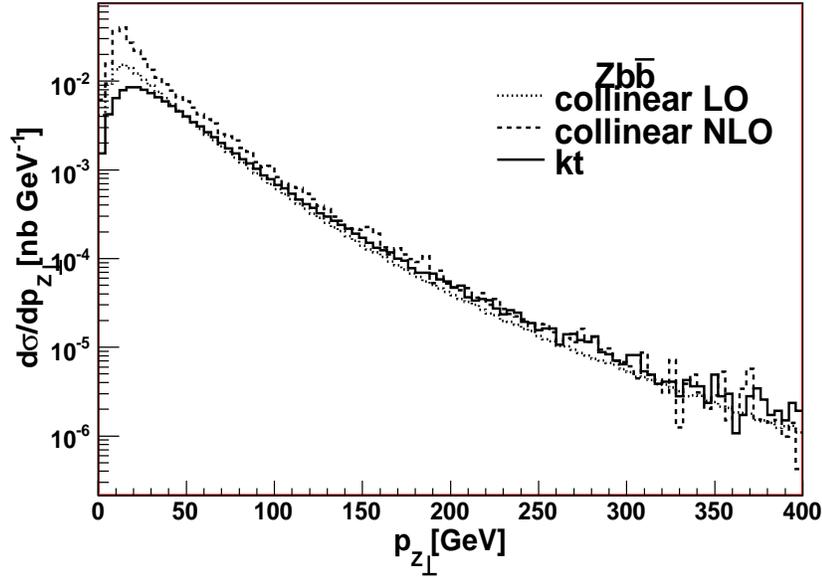,width=12cm,height=8.5cm,clip}
\caption{Comparison of cross sections differential in transverse
momentum of the produced $Z$ gauge boson.
Calculation with massless $b$-quarks. The applied cuts are described in
the text.}\label{fig:ZptNLO}
\end{center}
\end{figure}

\begin{figure}
\begin{center}
\epsfig{figure=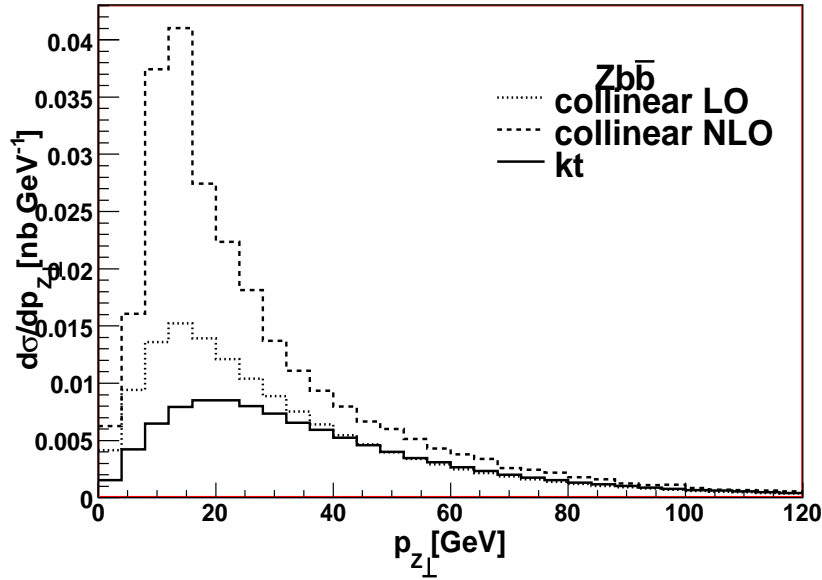,width=12cm,height=8.5cm,clip}
\caption{Comparison of cross sections differential in transverse
momentum of the produced $Z$ gauge boson (linear scale). Calculation
with massless $b$-quarks. The applied cuts are described in
the text.}\label{fig:ZptzNLO}
\end{center}
\end{figure}

The result for the cross sections differential in the transversal momentum
of $Z$ can be seen in Fig.~\ref{fig:ZptNLO}. The cross section
changes especially at small $p_{Z\perp}$ (see
Fig.~\ref{fig:ZptzNLO}) from LO to NLO calculation, and the
difference between collinear calculation and $k_T$-factorization
calculation becomes more pronounced. We observe that the maximum of the
distribution in the NLO calculation ({\sc Mcfm}) stays approximately
at same value of transversal momenta and the shape of the peak is
very different from the one we obtain in $k_T$-factorization.
Nevertheless, the $p_{Z\perp}$ distributions match at very high
$p_{Z\perp}$ ($\mathcal{O}(10^2 {\rm GeV})$).

\begin{figure}
\begin{center}
\epsfig{figure=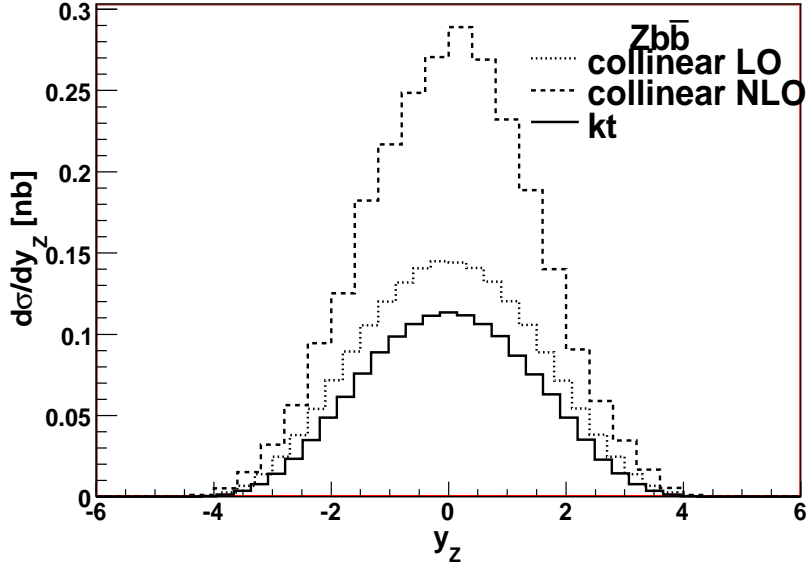,width=12cm,height=8.5cm,clip}
\caption{Comparison of cross sections differential in rapidity
of the produced $Z$ gauge boson (logarithmic scale).
Calculation with massless $b$-quarks. The applied cuts are described in
the text.}\label{fig:ZyNLO}
\end{center}
\end{figure}

The rapidity distribution of the $Z$ (Fig.~\ref{fig:ZyNLO}) shows
no major difference in shape in $k_T$-factorization approach, LO and NLO collinear
factorization approach.

\begin{figure}
\begin{center}
\epsfig{figure=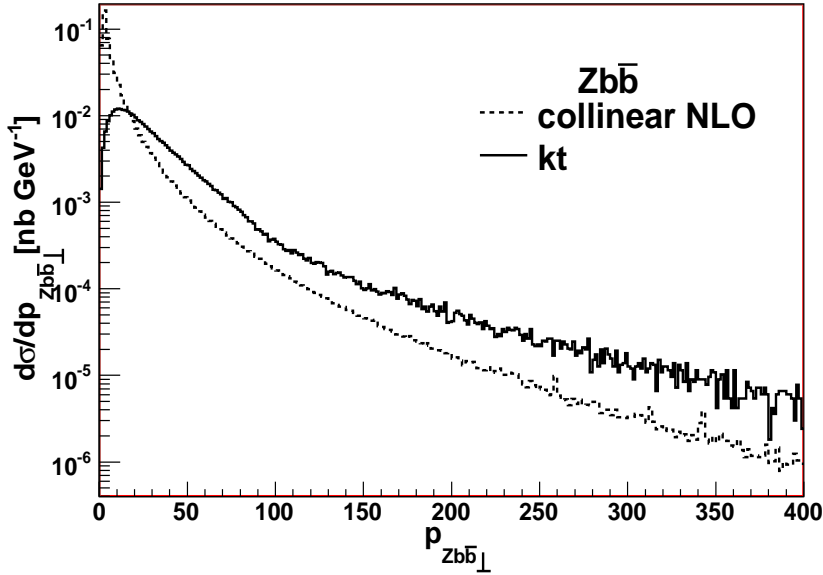,width=12cm,height=8.5cm,clip}
\caption{Comparison of cross sections differential in the
$p_{\perp}$ of the system $Zb\bar{b}$. Calculation with massless
$b$-quarks. The applied cuts are described in
the text.}\label{fig:ZbbptNLO}
\end{center}
\end{figure}

We consider the cross section differential in the total
transversal momentum of the $Zb\bar{b}$ system
$p_{Zb\bar{b}\perp}$ in Fig.~\ref{fig:ZbbptNLO}. In the NLO
collinear calculation  a non-zero $p_{Zb\bar{b}\perp}$ is generated by the
emission of an additional gluon, while at LO it is always balanced to
zero. At low $p_{Zb\bar{b}\perp}$ we see the consequence of the
cut on the transverse momenta of the outgoing particles in {\sc
Mcfm} (a small gap between $0~{\rm GeV}$ and $4.62~{\rm GeV}$ in $p_{Zb\bar{b}\perp}$ histogram). Since there are no parton showers or soft gluon
re-summation \cite{Balazs:1997xd,*Ellis:1997ii} included in the
{\sc Mcfm} NLO calculation, one observes a steep rise of the cross
section towards zero transverse momentum because the matrix
element diverges when approaching $p_{Zb\bar{b}\perp}\!\!\to
\!0\,{\rm GeV}$. On the other hand, uPDFs include corrections
similar to parton shower effects, treated consistently, which
causes the turnover in the cross section of the
$k_T$-factorization calculation. Here, the entire transversal
momentum of the $Zb\bar{b}$ system stems from the transversal
momenta of initial state gluons. 
We expect that resummation effects at
low values of $p_{Zb\bar{b}\perp}$ would tame the growth of the cross
section in collinear factorization and would decrease the difference to
$k_T$-factorization. 
Interestingly, there is a
difference not only at low values of $p_{Zb\bar{b}\perp}$, but
also at high values of $p_{Zb\bar{b}\perp}$. 
The differential cross sections at high  $p_{Zb\bar{b}\perp}$ have a
similar slope, but differ by a factor of $\sim 3$. This is contrary to
the behavior of distributions of $p_{Z\perp}$ in Figs.~\ref{fig:ZptNLO}
and \ref{histoZpt2} where at large values of $p_{Z\perp}$  the
differential cross sections overlap. 
For this difference at large $p_{Zb\bar{b}\perp}$
further calculations have to reveal the exact effect of higher order
corrections in collinear factorization, keeping in mind that the NLO for
this obersvable  de facto is the first non trivial
order.

\begin{figure}
\begin{center}
\epsfig{figure=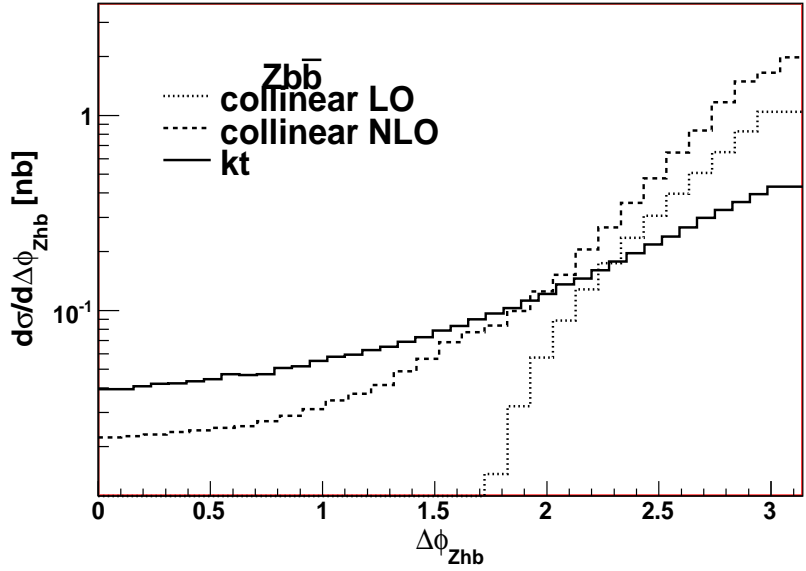,width=12cm,height=8.5cm,clip}
\caption{Comparison of cross sections differential in distance in
azimuthal angle of $Z$ and higher $p_\perp$ $b/\bar{b}$.
Calculation with massless $b$-quarks. The applied cuts are described in
the text.}\label{fig:angdNLO}
\end{center}
\end{figure}

\begin{figure}
\begin{center}
\epsfig{figure=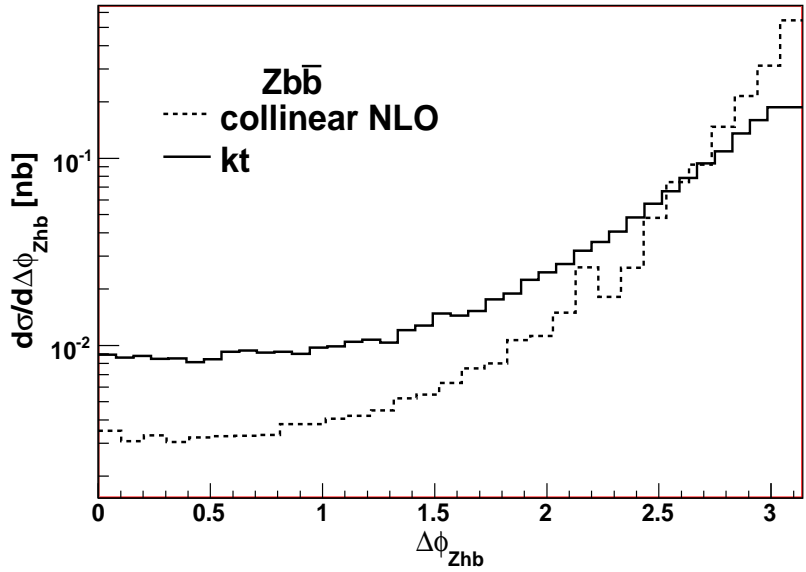,width=12cm,height=8.5cm,clip}
\caption{Comparison of cross sections differential in distance in
azimuthal angle of $Z$ and higher $p_\perp$ $b/\bar{b}$.
Calculation with massless $b$-quarks. An additional cut on $p_{Z\perp}>50\,{\rm GeV}$ has been applied.}
\label{fig:angdSteeper}
\end{center}
\end{figure}

The cross section differential in the difference of azimuthal
angles of $Z$ and $b$ or $\bar{b}$ quark with higher transversal
momentum -- $\Delta\phi_{Zhb}$ -- is shown in Fig.~\ref{fig:angdNLO}. Going from  LO to NLO, the
collinear calculation reveals a broader distribution like in the
$k_T$-factorization case. Nevertheless, the $k_T$-factorization
result shows a more homogeneous spread of the azimuthal angle
distance. This difference origins partly in the difference of the
transversal momentum distributions at low values (see
Fig.~\ref{fig:ZptzNLO}). A cut on low values
($p_{Z\perp}>50\,{\rm GeV}$) of the transversal momentum of the $Z$ boson
results in steeper $\Delta\phi_{Zhb}$ distributions as shown in 
Fig.~\ref{fig:angdSteeper}. 
Still, the $k_T$-factorization 
result is flatter than the 
NLO collinear factorization calculation giving an indication that there is a
contribution from the total transversal momentum of the $Zb\bar{b}$ system generated by
both uPDFs.

\subsection{Variation of the {\sc Cascade} results on uPDF and renormalization scale}\label{uPDFs}

\begin{figure}
\begin{center}
\epsfig{figure=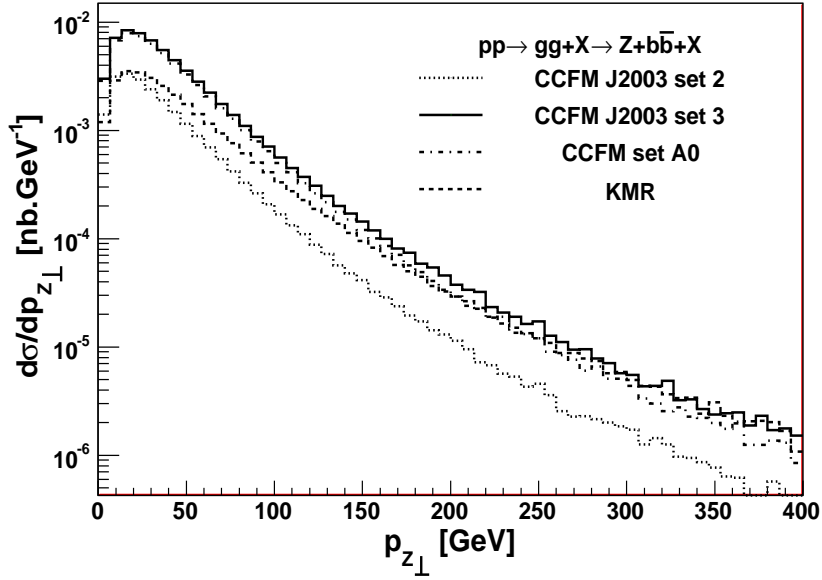,width=12cm,height=8.5cm,clip}
\caption{Transverse momentum distributions of produced $Z$ gauge
boson calculated in {\sc Cascade} using massive quarks. Cases with
different uPDFs compared.} \label{pdfsptZ}
\end{center}
\end{figure}

\begin{figure}
\begin{center}
\epsfig{figure=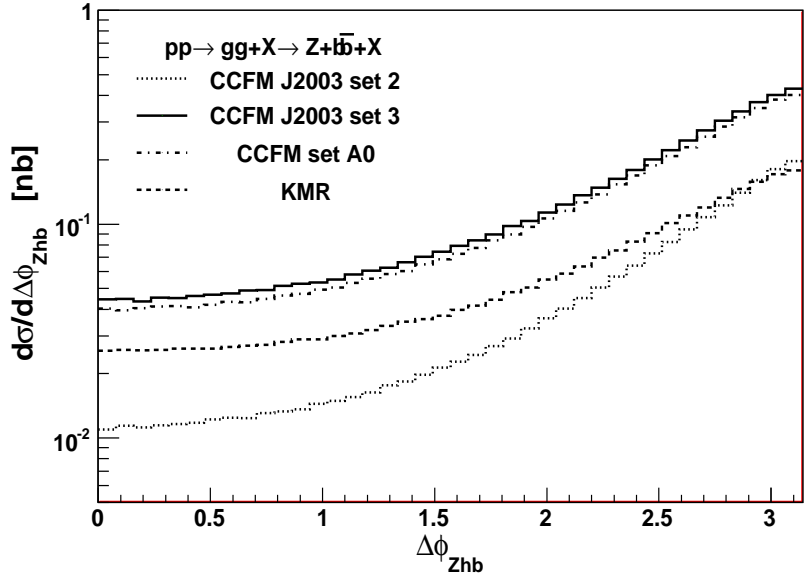,width=12cm,height=8.5cm,clip}
\caption{Comparison of cross sections differential in distance in
azimuthal angle of $Z$ and higher $p_\perp$ $b/\bar{b}$,  using massive quarks. Cases with
different uPDFs compared.} \label{pdfsphi}
\end{center}
\end{figure}

To estimate the uncertainty coming from the different choices of
uPDF sets, we
calculate the cross sections differential in either the transverse momentum
of the $Z$ boson or $\Delta\phi_{Zhb}$ (distance in polar angle
between $Z$ and max$(p_{b\perp},p_{\bar{b}\perp})$) using different
sets of uPDFs, namely CCFM J2003 set 1, 2, 3 \cite{Hansson:2003xz}
and CCFM set A0 \cite{Jung:2004gs}, which are all obtained from
fits to HERA $F_2$ data
\cite{Aid:1996au,*Adloff:2000qk,*Derrick:1996hn,*Chekanov:2001qu}.
In addition we use the unintegrated parton density by
\cite{Kimber:2001sc}, referred to as KMR. The resulting plots are
shown in Figs.~\ref{pdfsptZ} and \ref{pdfsphi}. We do not show the
distributions for set 1, because they are very close to
distribution for the set 3, to keep the plot clear.

The total cross sections obtained for different uPDFs can be seen
in Tab.~\ref{pdfscs}.
The total cross section varies for these different uPDFs about
$45\%$, while  the shape of the distributions is hardly effected
except of the KMR. KMR set uses completely different evolution
equations and a deviation is not surprising.

As a last point to discuss, we turn to  the scale dependence. 
As already mentioned in the beginning of 
section~\ref{sec:numstudies} the factorization scale is fixed by
the emission angle of the hard subprocess.  However, there is still
freedom in choice of the renormalization scale which should be of
order of the typical scale of the hard subprocess.

\begin{figure} \begin{center} 
\epsfig{figure=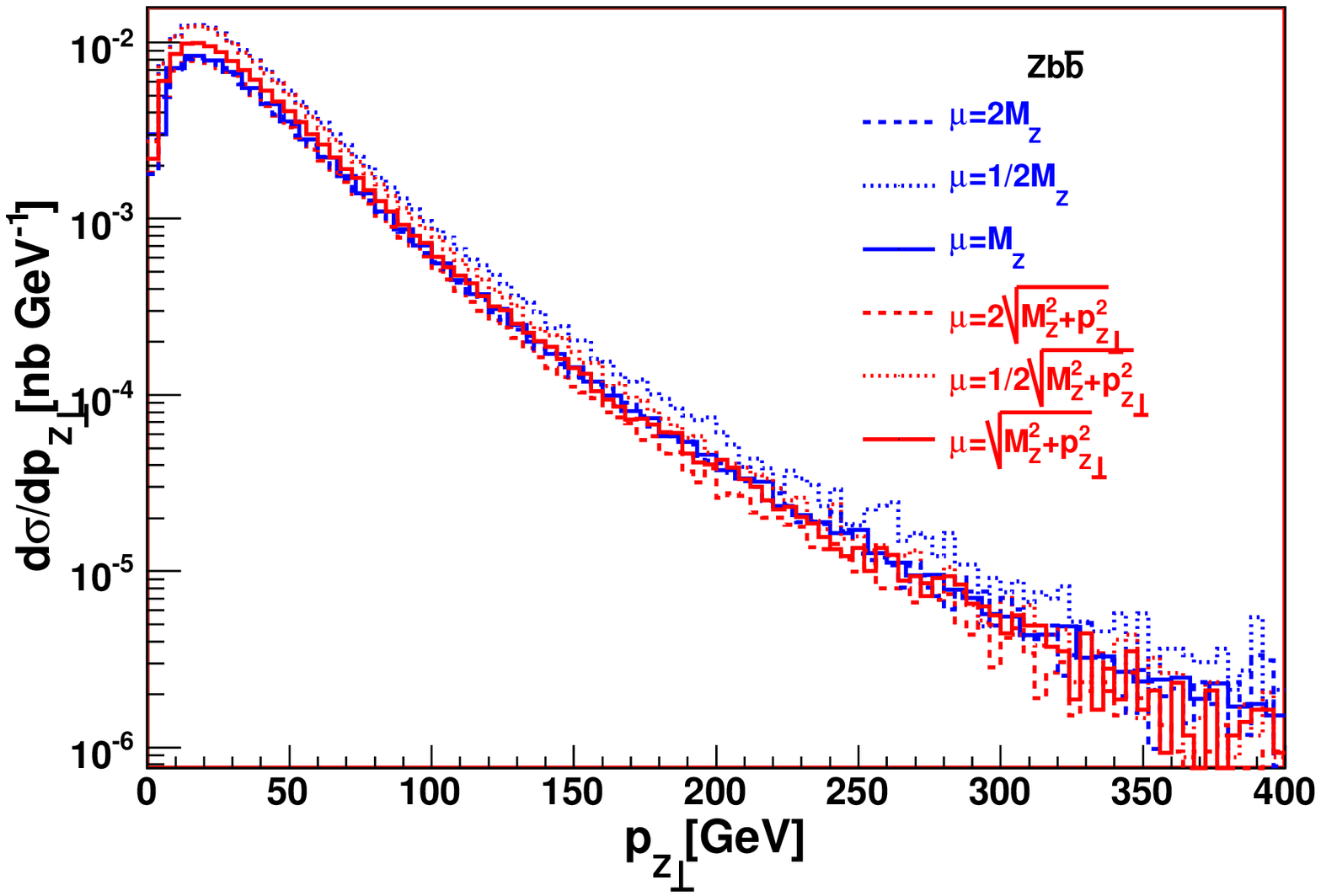,width=12cm,height=8.5cm,clip}
 \caption{Transverse momentum distributions of produced $Z$ gauge boson calculated in {\sc Cascade} using massive quarks. Cases with different renormalization scales $\mu_R$ compared.}\label{fig:ptScales}\end{center}\end{figure}

\begin{figure}\begin{center}
\epsfig{figure=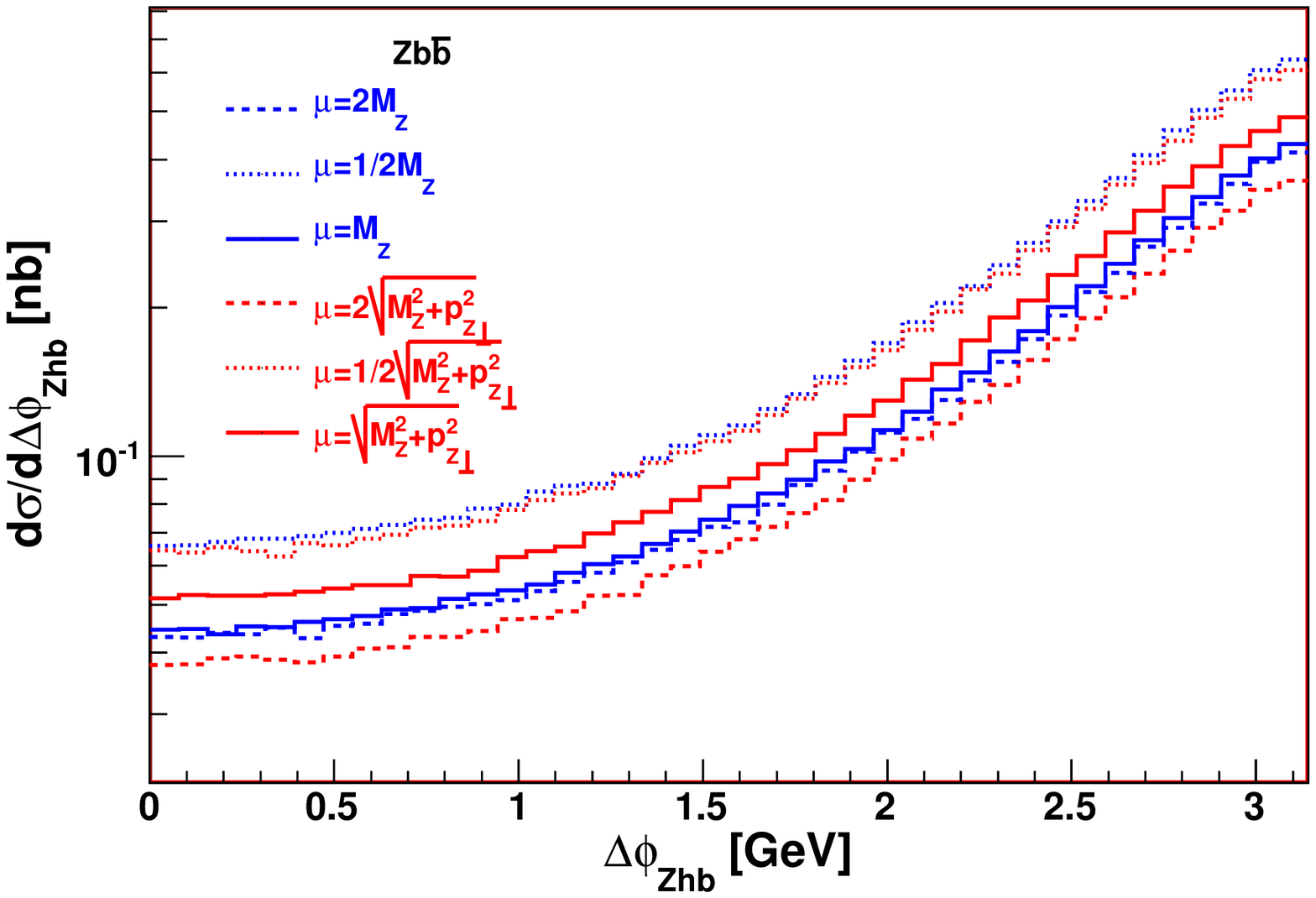,width=12cm,height=8.5cm,clip}
\caption{Transverse momentum distributions of produced $Z$ gauge boson calculated in {\sc Cascade} using massive quarks. Cases with different renormalization scales $\mu_R$ compared.}\label{fig:yZScales}\end{center}\end{figure}

We consider two possible choices: the constant renormalization scale $\mu_1=m_Z$
and the scale $\mu_2=\sqrt{m_Z^2+p_{Z\perp}^2}$, which are varied
by factor of $2$, so $\mu$ has values $2\mu_1$, $\frac{1}{2}\mu_1$
and $2\mu_2$, $\frac{1}{2}\mu_2$. The results for the $p_{Z\perp}$
and the $\Delta\phi_{Zhb}$ distribution can be seen in
Figs.~\ref{fig:ptScales} and \ref{fig:yZScales}, respectively. The
values of the cross section for individual choices of the scale
are summarized in Tab.~\ref{scalescs}. One can see that a running $\alpha_S$ does not affect the
shape of the distributions, but only the total cross section.

\begin{table}
\begin{center}
\begin{tabular}{|c|c|}
\hline
uPDF & Total cross section [nb]\\
\hline
      CCFM J2003 set 1        &   $0.369$               \\
      CCFM J2003 set 2        &   $0.147$               \\
      CCFM J2003 set 3        &   $0.406$               \\
      CCFM set B0       &   $0.277$               \\
      CCFM set A0       &   $0.378$               \\
      KMR      &   $0.190$               \\
\hline
\end{tabular}
\end{center}
\caption{Total cross sections of the process $pp\to Zb\bar{b}+X$ for different sets of unintegrated
parton distribution functions.}\label{pdfscs}
\end{table}

\begin{table}
\begin{center}
\begin{tabular}{|c|c|}
\hline
$\mu_R$ & Total cross section [nb]\\
\hline
      $m_Z$        &   $0.406$               \\
      $2m_Z$       &   $0.392$               \\
$\frac{1}{2}m_Z$       &   $0.607$               \\
      $\sqrt{m_Z^2+p_{Z\perp}^2}$       &   $0.467$               \\
      $2\sqrt{m_Z^2+p_{Z\perp}^2}$       &   $0.381$               \\
$\frac{1}{2}\sqrt{m_Z^2+p_{Z\perp}^2}$      &   $0.585$               \\
\hline
\end{tabular}
\end{center}
\caption{Total cross sections for different renormalization scale
$\mu$.}\label{scalescs}
\end{table}

\section{Summary and Conclusions}

In this paper we have calculated the matrix element for the process
$g^*g^*\rightarrow W/Z q_i\bar{q}_j$, taking into account the virtuality and
transversal momenta of the initial gluons in the $k_T$-factorization formalism.
We have implemented the matrix element squared in the Monte Carlo generator {\sc Cascade} and have calculated
the total and differential cross sections of this process in proton proton collisions for the LHC at energy of $\sqrt{s}=14{\rm TeV}$. We have compared our
results with results obtained in collinear factorization (using {\sc Mcfm}).
The total cross sections differ by a factor of $\sim2$. 
There are differences in distributions which are sensitive to
compensation of transversal momenta of particles in the final
state coming from rather fundamental differences between the two approaches.

We found the most significant differences in the cross section differential 
in the azimuthal angle between the $Z$ boson and higher $p_T$ quark or 
antiquark -- $\Delta\phi_{Zhb}$. While for a LO calculation in collinear factorization 
a region of values of $\Delta\phi_{Zhb}$ is kinematically forbidden, in $k_T$-factorization 
the whole range of $\Delta\phi_{Zhb}$ is allowed. This is because of neglecting the 
contribution of transversal momenta of initial state gluons in calculation of matrix
element in collinear factorization.  
The  NLO collinear calculation (where transversal momentum is generated by real 
corrections) shows already the same qualitative behavior as the $k_T$-factorization  calculation.
However, there remains a difference in the
shape of the distribution of $\Delta\phi_{Zhb}$ compared to the $k_T$-factorization calculation 
We also compared cross sections differential in the transversal momentum of the $Zb\bar{b}$ system -- $p_{Zb\bar{b}\perp}$.
In collinear factorization and  lowest order perturbation theory ($\alpha_S^2$),
the observable $p_{Zb\bar{b}\perp}$ is exactly zero. For a non-zero contribution in collinear factorization higher order corrections are needed.
The $k_T$-factorization gives non-zero
contribution already at $\alpha_S^2$ order. 
We have compared cross sections differential in $p_{Zb\bar{b}\perp}$ calculated in NLO in collinear calculation and LO in $k_T$-factorization. 
The distributions have different shape at low values of $p_{Zb\bar{b}\perp}$. At high $p_{Zb\bar{b}\perp}$ the slopes are similar but differ in absolut size.

We have calculated the cross sections differential in the transversal momentum of the produced boson.
The maximum of the distribution in the $k_T$-factorization calculation is at higher
transversal momenta compared to the collinear one. This shows the sensitivity of this
distribution on parton evolution model and treatment of kinematics.

We conclude that 
some of the effects of NLO and even higher order collinear calculation are already
included in the LO $k_T$-factorization calculation.

\vspace{.5cm}\noindent {\bf Acknowledgments}

\noindent

First of all, we would like to thank H. Jung and J. Bartels for their help, advice, and support during all stages of this work.
We would like to express special thanks to S.~P. Baranov
for useful advice and introduction to method of orthogonal
amplitudes, which was crucial to obtain numerically stable results.
We are thankful to J. Campbell for help with filling of histograms
in Monte Carlo generator {\sc Mcfm}.
We also benefited from
interesting discussions with A. Bacchetta. F.S. has been supported
in part by the Deutsche Forschungsgemeinschaft, DFG Grant No. GRK 602,
and the Agence Nationale de la Recherche (France), contract No ANR-06-JCJC-0084-02.
This work is associated with the DFG Collaborative Research Centre SFB 676.

\providecommand{\etal}{et al.\xspace}
\providecommand{\href}[2]{#2}
\providecommand{\coll}{Coll.}
\catcode`\@=11
\def\@bibitem#1{%
\ifmc@bstsupport
  \mc@iftail{#1}%
    {;\newline\ignorespaces}%
    {\ifmc@first\else.\fi\orig@bibitem{#1}}
  \mc@firstfalse
\else
  \mc@iftail{#1}%
    {\ignorespaces}%
    {\orig@bibitem{#1}}%
\fi}%
\catcode`\@=12

\end{document}